\documentclass[aps,prl,twocolumn,groupedaddress]{revtex4-1}
\usepackage{amsmath}
\usepackage{amssymb}
\usepackage{graphicx}
\usepackage{dcolumn}
\usepackage{bm}

\begin{document}

\title{Anisotropic metallic metasurface superlattices supporting Fano resonances and bound states in the continuum}

\author{Sun-Goo Lee}
\email{sungooleee@gmail.com}
\author{Seong-Han Kim}
\author{Chul-Sik Kee}
\email{cskee@gist.ac.kr}
\affiliation{Integrated Optics Laboratory, Advanced Photonics Research Institute, GIST, Gwangju 61005, South Korea}
\date{\today}

\begin{abstract}
A perfect metal film with a periodic arrangement of cut-through slits, an anisotropic metallic metamaterial film, mimics a dielectric slab and supports guided electromagnetic waves in the direction perpendicular to the slits. Here, we introduce metallic metasurface superlattices that include multiple slits in a period, and demonstrate that the superlattices support the Fano resonances and bound states in the continuum. The number of Fano resonances and bound states depend on the number of slits in a period of superlattices. The metallic metasurface superlattices provide new mechanisms to manipulate electromagnetic waves, ranging from microwave to far-infrared wavelengths, where a conventional metal can be considered as a perfect electric conductor.
\end{abstract}

\pacs{78.67-n, 42.70.Qs}

\maketitle

Manipulating electromagnetic waves by utilizing the planar subwavelength metastructures that support the quasiguided Bloch modes, including one-dimensional (1D) gratings and two-dimensional (2D) photonic crystal slabs, is of fundamental importance in photonics. While a conventional layer of naturally existing dielectric material obeys the Fresnel equations and Snell's law \cite{Born2002}, a periodic metasturcture film can capture the incident light owing to the lateral Bloch modes and reemit the captured light resonantly \cite{YHKo2018}. With proper Fano or guided-mode resonances, diverse unusual spectral responses, which cannot be achieved from naturally occurring materials, can be realized in a significantly compact format, even as a single layer film. Various optical devices, such as filters \cite{WShu2003,MNiraula2015,Kawanishi2020}, reflectors \cite{Magnusson2014,JWYoon2015}, polarizers \cite{KLee2014,Hemmati2019}, sensors \cite{YLiu2017,Abdallah2019}, and lasers \cite{Kodigala2017,STHa2018} have been realized based on the resonance effects.

In planar photonic lattices, the Fano resonances arise because the quasiguided Bloch modes with finite $Q$ factors resonantly exchange the electromagnetic energy with the radiating waves in the radiation continuum. In some circumstances, however, the Bloch mode in the radiation continuum is completely decoupled from the radiating waves and becomes a bound state in the continuum (BIC), which has lately garnered considerable scientific interest \cite{Marinica2008,Plotnik2011,Hsu2016,Minkov2018,Koshelev2019}. BICs are associated with diverse interesting physical phenomena, such as high-$Q$ Fano resonances \cite{Koshelev2018,Abujetas2019}, enhanced nonlinear effects \cite{Koshelev2020}, and topological natures \cite{BZhen2014,Doeleman2018,SGLee2019-1}. Different types of BICs have been studied in versatile photonic lattices \cite{Yang2014,JJin2019,XYin2020,Bulgakov2018,XGao2019}. 

In the past few decades, the utilization of periodic subwavelength metallic structures to create artificial materials with unusual electromagnetic properties was significantly investigated \cite{Pendry2004,JTShen2005,CRWilliams2008,JShin2009}. It was shown that a perfect metal film with a 1D subwavelength arrangement of cut-through slits, i.e., an anisotropic metallic metamaterial film, can be considered as a dielectric slab and support the guided modes propagating along the film \cite{JTShen2005}. The realization of Fano resonances and BICs by utilizing perfect metal films is an interesting research topic because it could provide new mechanisms to manipulate the electromagnetic waves, from microwave to far-infrared wavelengths. However, to the best of our knowledge, no study has been conducted on the Fano resonance and BIC in anisotropic metallic metamaterial films thus far. In this study, we first examine why the conventional anisotropic metallic metamaterial films do not exhibit the leaky-wave effects, and present anisotropic metallic metasurface superlattices supporting the Fano resonances and BICs.

\begin{figure}[t]
\includegraphics[width=8.3cm]{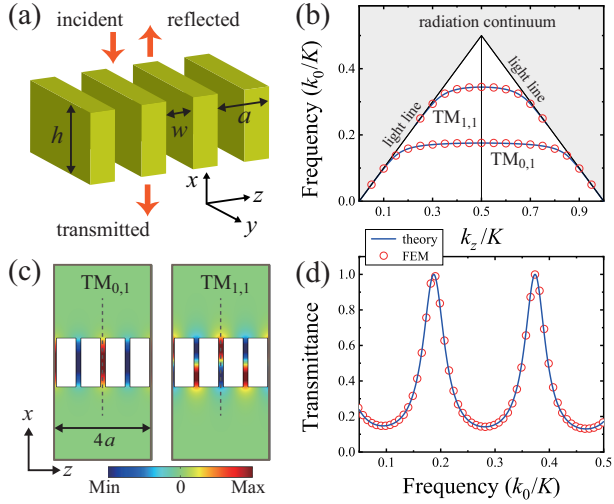}
\caption {\label{fig1} (a) Schematic of a metallic metamaterial film. (b) Photonic bands for the lowest even ($\mathrm{TM}_{0}$) and odd ($\mathrm{TM}_{1}$) modes. (c) Spatial magnetic field ($H_y$) distributions of the $\mathrm{TM}_{0}$ and $\mathrm{TM}_{1}$ modes at $k_z =0.5~K$. White regions indicate perfect metals. Band edge modes have symmetric field distributions according to the mirror planes represented by dotted lines. (d) Zeroth order transmittance through the metamaterial film. Solid lines and open circles are obtained from the diffraction theory and FEM simulations, respectively.  }
\end{figure}

Figure~\ref{fig1}(a) illustrates a conventional metallic metamaterial film with one slit in period $a$. The thickness of the film is $h$ and width of the air slit is $w$. The ambient and slit regions are assumed to be air and the metallic parts are assumed to be perfect electric conductors. Generally, the metamaterial film supports multiple transverse magnetic ($\mathrm{TM}_{q}$) modes propagating along the $z$ direction; moreover, by  the analytical diffraction theory \cite{JTShen2005}, the even ($q=0, 2, 4, \cdots $) and odd ($q=1, 3, 5, \cdots $) modes satisfy the dispersion relations
\begin{equation}\label{even}
\phi =\cot{\left ( \frac{k_0 h}{2} \right )}
\end{equation} 	 	
and
\begin{equation}\label{odd}
\phi =-\tan{\left ( \frac{k_0 h}{2} \right )},
\end{equation} 	 	
respectively, where
\begin{equation}\label{phi}
\phi =\sum_{p=-\infty}^{\infty}  \frac{f k_0}{\sqrt{G_p^2 - {k_0}^2}}\left [  \frac{\sin(G_p w/2)}{G_p w/2} \right]^2
\end{equation} 	 	
with $f=w/a$ and $G_p=k_z + pK$. Here, $K=2\pi/a$ is the magnitude of the grating vector and $p$ is an integer representing the diffraction order. The zeroth order transmission amplitude in the subwavelength regime is given by
\begin{equation}\label{transmission}
t_0 = \frac{4[(f/\phi^2)/(1+1/\phi)^2]e^{ik_0 h}}{1-[(1-1/\phi)/(1+1/\phi)]^2e^{2ik_0 h}}
\end{equation} 	 	
from the diffraction theory. Figure~\ref{fig1}(b) represents the photonic bands of the lowest even ($q=0$) and odd ($q=1$) modes calculated by the diffraction theory (solid lines) and finite element method (FEM) simulations (open circles), and Fig.~\ref{fig1}(c) illustrates the corresponding spatial magnetic field ($H_y$) distributions at $k_z=0.5~K$, obtained by the FEM simulations when $h=2.5~a$ and $w=0.2~a$. While the $\mathrm{TM}_{q}$ modes in ordinary dielectric photonic crystal slabs exhibit multiple photonic bands, $\mathrm{TM}_{q,n}$ ($n=1, 2, 3, \cdots $), and band gaps, $\mathrm{\Delta}_{q,n}$, between the $\mathrm{TM}_{q,n}$ and $\mathrm{TM}_{q,n+1}$ bands in the radiation continuum as well as below the light line, as evident in Fig.~\ref{fig1}(b), the anisotropic metamaterial films allow only one band, $\mathrm{TM}_{q,1}$, for each $\mathrm{TM}_{q}$ mode below the light line, and there is no band gap. We note that Eqs.~(\ref{even}) and (\ref{odd}) cannot be satisfied in the gray region of the radiation continuum because the $\phi$ in Eq.~(\ref{phi}) has complex values above the light line. The guided modes in the $\mathrm{TM}_{q,1}$ bands below the light line are completely decoupled from the radiating waves by total internal reflection; thus, there is no lateral resonance effect in the transmittance curve through the metamaterial film, as shown in Fig.~\ref{fig1}(d). To realize a Fano resonance and BIC, therefore, higher-order $\mathrm{TM}_{q,n \geq 2}$ bands in the radiation continuum are essential.

The reason why the metallic metamaterial film, shown in Fig.~\ref{fig1}(a), allows only one band per mode without a photonic band gap can be understood from the origin of the photonic band gap with the concept of ``dielectric band'' and ``air band'' in the photonic band structures of a representative 1D photonic crystal that consists of alternating layers of dielectric materials. Conventionally, the dielectric (air) band indicates the photonic band below (above) the first band gap that opens at $k_z =0.5~K$.  According to the electromagnetic variational theorem, photonic band gaps arise because the modes in the dielectric (air) band tend to locate their field energy in the regions of a high (low) dielectric constant \cite{Joannopoulos1995}. One band edge mode has symmetric field distributions, while the other one has asymmetric distributions with respect to the mirror plane of symmetry because the band edge modes at the Brillouin zone boundaries should be standing waves. In metallic metamaterial films, however, the field energy should be distributed only in the narrow subwavelength slit regions because the electromagnetic waves cannot exist inside the perfect electric conductors, as shown in Fig.~\ref{fig1}(c). Hence, neither a higher-order $\mathrm{TM}_{q,n \geq 2}$ band nor a photonic band gap exists in the perfect metal films with cut-through slits.

We now introduce and analyze metallic metasurface superlattices that can support higher-order photonic bands and band gaps. In the proposed concept, it is important to increase the degree of freedom to distribute the electromagnetic energy by increasing the number of slits in a period of superlattices. The photonic band structures and transmission properties of metasurface superlattices are investigated by varying the slit parameters, such as the numbers, widths, and positions, through FEM simulations. We note that the rigorous FEM simulations sufficiently describe the dispersion and transmission properties of the photonic systems composed of perfect metals, as shown in Fig.~\ref{fig1}. The thickness is set to $h=1.5~a$ and we limit our attention to the lowest even $\mathrm{TM}_{0}$ mode below $k_0/K=0.5$, because this simple case can sufficiently demonstrate the key properties of the metallic metasurface superlattices.

\begin{figure}[t]
\includegraphics[width=8.3cm]{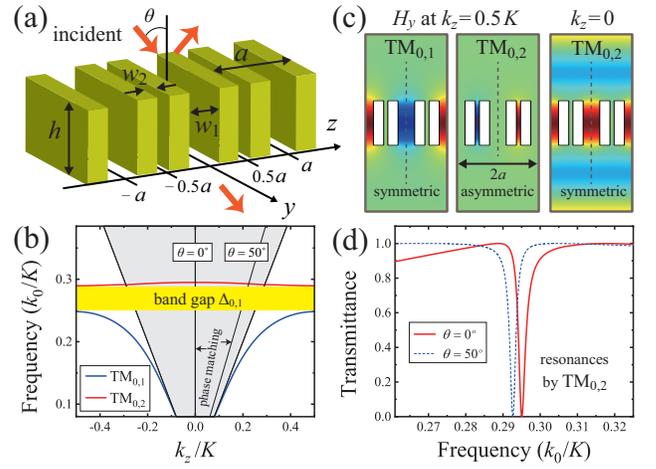}
\caption {\label{fig2} (a) Metallic metasurface superlattice with two slits in a period. (b) Simulated photonic bands for the lowest even $\mathrm{TM}_{0}$ mode. (c) Spatial magnetic field distributions of Bloch modes at $k_z =0$ and $0.5~K$. (d) Zeroth order transmittance through the metallic metasurface superlattice. }
\end{figure}

Figure~\ref{fig2}(a) shows the simplest metallic metasurface superlattice that contains two slits with different widths in period $a$. Slits with width $w_1=0.4~a$ are located at $z=0, \pm a, \pm 2~a, \cdots$ and those with $w_2=0.1~a$ are located at $z=\pm 0.5~a, \pm 1.5~a, \cdots$. The simulated dispersion relations illustrated in Fig.~\ref{fig2}(b) show that the superlattice with two slits supports two photonic bands, $\mathrm{TM}_{0,1}$ and $\mathrm{TM}_{0,2}$, and a photonic band gap, $\mathrm{\Delta}_{0,1}$, between the two bands. While the $\mathrm{TM}_{0,1}$ band lies only in the white region below the light line, the $\mathrm{TM}_{0,2}$ band exists in the gray region of the radiation continuum as well as below the light line. The spatial magnetic field distributions plotted in Fig.~\ref{fig2}(c) show that the band edge modes at $k_z =0.5~K$ in the $\mathrm{TM}_{0,1}$ and $\mathrm{TM}_{0,2}$ bands are well confined in the metasurface because they are protected by the total internal reflection, but a Bloch mode at $k_z =0$ in the $\mathrm{TM}_{0,2}$ band is radiative out of the metasurface because it is in the radiation continuum. The radiative Bloch modes in the $\mathrm{TM}_{0,2}$ band generate Fano resonances with the phase matching condition $k_0 \sin \theta =k_z$ \cite{YDing2007} in the transmission curves, as illustrated in Fig.~\ref{fig2}(d). Because the $\mathrm{TM}_{0,2}$ band is significantly flat in the grey region of the radiation continuum, the position of resonances moves slightly in the transmission spectra with variations in the incident angle $\theta$.

At $k_z =0$, as shown in Fig.~\ref{fig2}(c), the Bloch mode has symmetric field distributions with a period of $a$. At $k_z =0.5~K$, in contrast, the band edge modes in the lower $\mathrm{TM}_{0,1}$ and upper $\mathrm{TM}_{0,2}$ bands have symmetric and asymmetric field distributions, respectively, with a period of $2~a$. In conventional 1D dielectric photonic crystals, the band edge modes of the $n$th band gap, $\mathrm{\Delta}_{q,n}$, should be symmetric or asymmetric standing waves with a period of $a\times 2^{(2-n)}$. Based on the two constraints on the period of standing waves and field distributions in the slit regions, in metallic metasurface superlattices, we conjecture that the number of degree-of-freedom-allocating field distributions at $k_z =0.5~K$ is the same as the number of slits in a period of superlattices. This concept can then be extended to generic $k_z$ points. With our conjecture, metallic metasurfaces with $N$ slits in a period can allow $N$ photonic bands and $N-1$ band gaps for individual $\mathrm{TM}_{q}$ modes. Because the symmetry-protected BICs generally appear at the edges of the second stop bands in diverse periodic photonic structures, it is to be expected that they will also be found in metasurface superlattices with either three or more slits in a period.

\begin{figure}[b]
\centering
\includegraphics[width=8.3cm]{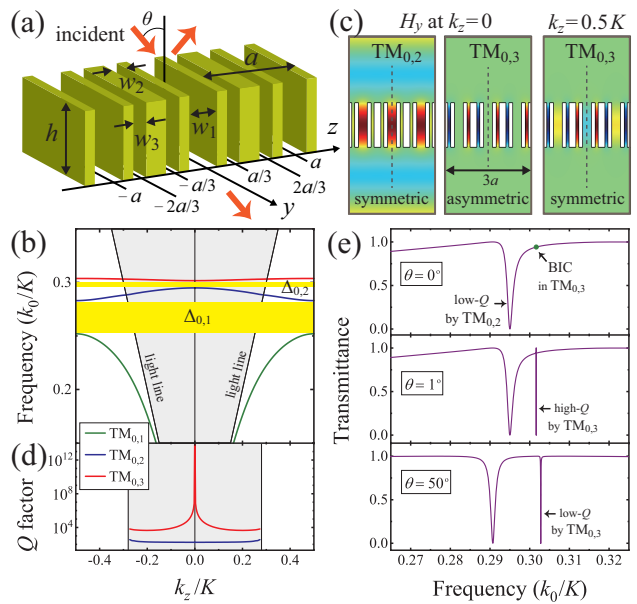}
\caption{\label{fig3} (a) Metallic metasurface superlattice with three slits in a period. (b) Simulated photonic bands for the lowest even $\mathrm{TM}_{0}$ mode. (c) Spatial magnetic field distributions of standing wave Bloch modes at $k_z =0$ and $0.5~K$. (d) Radiative $Q$ factors of the $\mathrm{TM}_{0,1}$ and $\mathrm{TM}_{0,2}$ bands in the radiation continuum. (e) Zeroth order transmittance at three different incident angles. }
\end{figure}

Figure~\ref{fig3}(a) illustrates a metasurface superlattice possessing three slits in period $a$. In principle, even though there are multiple ways to select the position and width of slits in a period, for simplicity, the slits with width $w_1=0.3~a$ are located at $z=0, \pm a, \pm 2~a, \cdots$ and those with $w_2=0.1~a$ ($w_3=0.1~a$) are located at $z=\pm a/3$ ($\pm 2~a/3$), $\pm 4~a/3$ ($\pm 5~a/3$), $\cdots$. As evident from the dispersion relations represented in Fig.~\ref{fig3}(b), the metasurface superlattice with three slits in a period supports three photonic bands, i.e., $\mathrm{TM}_{0,1}$, $\mathrm{TM}_{0,2}$, and $\mathrm{TM}_{0,3}$, and two band gaps, $\mathrm{\Delta}_{0,n}$, between the adjacent $\mathrm{TM}_{0,n}$ and $\mathrm{TM}_{0,n+1}$ bands. The simulated spatial magnetic field distributions shown in Fig.~\ref{fig3}(c) reveal that at $k_z =0$, the edge mode in the $\mathrm{TM}_{0,2}$ and $\mathrm{TM}_{0,3}$ bands has symmetric and asymmetric field distributions, respectively, with a period of $a$; moreover, at $k_z =0.5~K$, the Bloch mode in the $\mathrm{TM}_{0,3}$ band has symmetric field distributions with a period of $a/2$. This occurs because the period of the standing waves at the edges of the $n$th band gap, $\mathrm{\Delta}_{0,n}$, is given by $a\times 2^{(2-n)}$. At $k_z =0$, while the symmetric mode in the $\mathrm{TM}_{0,2}$ band is radiative out of the metasurface, the asymmetric mode in the $\mathrm{TM}_{0,3}$ band becomes a nonradiative symmetry-protected BIC. The existence of BICs can be elucidated by investing radiative $Q$ factors in the grey region of the radiation continuum depicted in Fig.~\ref{fig3}(d). In the $\mathrm{TM}_{0,3}$ band, the symmetry-protected BIC exhibits a $Q$ factor that is larger that $10^{15}$ at $k_z=0$, but the $Q$ values decrease abruptly and become less than $10^4$ as $k_z$ drifts from zero. In the $\mathrm{TM}_{0,2}$ band, the $Q$ values in the radiation continuum are less than $10^3$ and there are no abrupt changes in the $Q$ factors. An embedded BIC can also be verified in the transmission spectra represented in Fig.~\ref{fig3}(d). At normal incidence with $\theta = 0^\circ$, the metasurface with three slits exhibits a low-$Q$ resonance due to the leaky mode in the $\mathrm{TM}_{0,2}$ band, and the embedded BIC in the $\mathrm{TM}_{0,3}$ band does not generate the resonance effect in the transmission spectra. In contrast, as $\theta$ increases from zero, two resonances due to the $\mathrm{TM}_{0,2}$ and $\mathrm{TM}_{0,3}$ bands simultaneously appear in the transmission spectra and their positions follow the phase matching condition $k_0 \sin \theta =k_z$.

\begin{figure}[t]
\centering
\includegraphics[width=8.3cm]{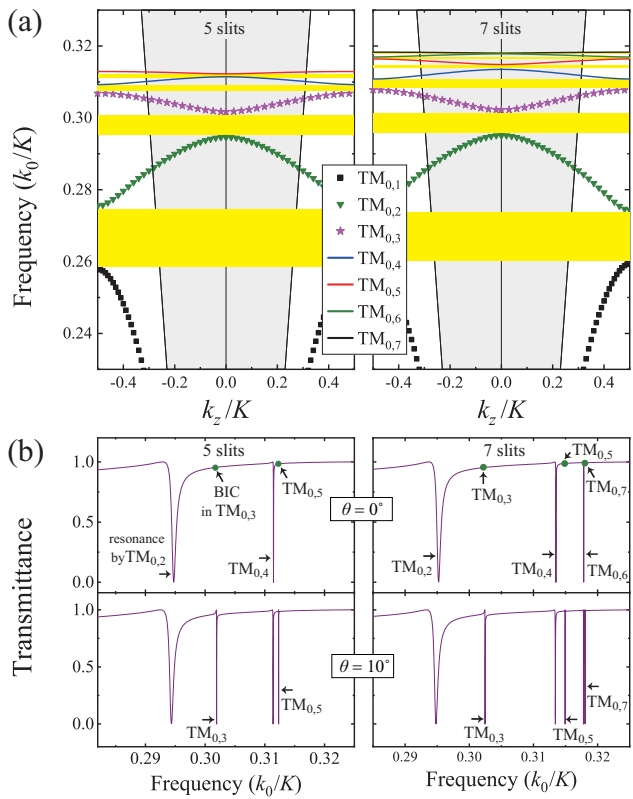}
\caption{\label{fig4} (a) Simulated photonic bands $\mathrm{TM}_{0,n}$ when there are 5 and 7 slits in a period. (b) Zeroth order transmittance through the metallic metasurface superlattice with an incident angle of $0^{\circ}$ and $10^{\circ}$. In the simulations, the slit widths are set as $w_c =0.18~a $ ($0.14~a$) and $w =0.08~a$ ($0.06~a$) when the number of slits is 5 (7). }
\end{figure}

The dispersion relations and transmission properties of metasurface superlattices were intensively investigated herein by varying the number of slits in a period, and it was numerically verified that $N$ slits in a period of superlattices allow $N$ photonic bands and $N-1$ band gaps for the $\mathrm{TM}_{0}$ mode. As examples, the photonic band structures of the metasurface superlattices with 5 and 7 slits, illustrated in Fig.~\ref{fig4}(a), evidently show that 5 (7) slits in a period exhibit 5 (7) photonic bands and 4 (6) band gaps. As demonstrated by the transmittance curves plotted in Fig.~\ref{fig4}(b), at normal incidence, two (three) Fano resonances are observed when the number of slits is 5 (7), but when $\theta = 10^{\circ}$, 4 (6) resonances are observed. In the simulations, the slits are located at $k_z=(m/N)\times a$, where $m$ is an integer and $N$ denotes the number of slits. The width of the slits at $k_z=0, \pm~a, \pm 2~a, \cdots$ is $w_c$, while that of others is $w\neq w_c$. Because the metasurface superlattices considered herein possess in-plane mirror symmetry with respect to the center of slits with width $w_c$, one of the two edge modes of band gap $\mathrm{\Delta}_{q,2n}$ that opens at $k_z=0$ becomes a symmetry-protected BIC (green solid circles) and the other mode with finite $Q$ factors generates a Fano resonance in the transmission curves. When $\theta \neq 0^{\circ}$, therefore, the number of resonances is equal to the number of  bands ($N-1$) in the radiation continuum. However, when $\theta = 0^{\circ}$, the symmetry-protected BICs are not shown in the transmission spectra.

In conclusion, we introduced metallic metasurface superlattices that contain multiple slits in a period, and demonstrated that the metasurface superlattices support the Fano resonances and BICs. Because the number of slits in a period is equal to the number of degree-of-freedom-distributing electromagnetic energies in the slit regions, metasurface superlattices with $N$ slits can support $N$ photonic bands and $N-1$ band gaps. With nonzero incident angles, $N-1$ resonances are observed in the transmission spectra due to the $N-1$ phase-matched Bloch modes. With normal incidence, in contrast, the symmetry-protected BICs at the edges of $\mathrm{\Delta}_{q,2n}$ band gaps are not shown in the transmittance curves. Our study is limited here to the lowest $\mathrm{TM}_{0}$ mode in 1D metastructures. However, the extension of this work to higher order $\mathrm{TM}_{q \geq  2}$ modes and 2D superlattices is feasible. By appropriately designing the slit parameters, such as the numbers, widths, and positions, the metasurface superlattices can be applied to generate diverse spectral responses in various wavelengths, ranging from microwave to far infrared, where a conventional metal can be considered to be a perfect electric conductor.

This research was supported by a grant from the National Research Foundation of Korea, funded by the Ministry of Education (No. 2020R1I1A1A01073945) and Ministry of Science and ICT (No. 2020R1F1A1050227), along with the Gwangju Institute of Science and Technology Research Institute in 2020.

\end{document}